\documentstyle[prl,aps]{revtex}
\newcommand{\eq}{\begin{equation}}
\newcommand{\en}{\end{equation}\noindent}
\newcommand{\eqa}{\begin{eqnarray}}
\newcommand{\ena}{\end{eqnarray}\noindent}

\def\log{{\rm log}}
\begin{document}
\input{epsf}
\draft
\title{Branched Growth with $\eta \approx 4$ Walkers}
\author{Thomas C. Halsey}
\address{Corporate Strategic Research, ExxonMobil Research and
Engineering, Route 22 East, Annandale, New Jersey 08801,
tchalse@erenj.com
}
\date{May 2, 2001}
\maketitle
\begin{abstract}
Diffusion-limited aggregation has a natural generalization to the
``$\eta$-models", in which $\eta$ random walkers must arrive at a point on the
cluster surface in order for growth to occur. It has recently been proposed that in
spatial dimensionality $d=2$, there is an upper critical $\eta_c=4$ above which the
fractal dimensionality of the clusters is $D=1$. I compute the first order
correction to $D$ for $\eta <4$, obtaining $D=1+\frac{1}{2}(4-\eta)$. The methods
used can also determine multifractal dimensions to first order in $4-\eta$. 
\end{abstract}

The formation of patterns in nature is often controlled by diffusive
phenomena. The branching of a less viscous fluid such as water injected
into a more viscous fluid such as oil, the dendritic complexity of a
snowflake, or the formation of veins of metals on the surface of certain
rocks, all display pattern formation processes controlled by the
diffusive transport of some quantity\cite{phystoday}.

The simplest model for diffusion-controlled growth was introduced twenty
years ago by Witten and Sander\cite{wittensander}. Their model of ``diffusion-limited
aggregation" (DLA) describes the formation of an aggregate by sequential
deposition of randomly walking particles arriving from infinity. There is
an electrostatic formulation of this process, in which the $n$-particle
cluster is chosen as an equipotential of a Laplacian field, which has a
source at infinity. The local growth probability, i.e., the probability
of deposition of the $n+1$'st particle, is then chosen proportional to
the local electric field on the surface of the cluster.

A natural generalization of this model is to fix the growth probability
on the surface proportional to the $\eta$'th power of the local electric
field.  This corresponds to a random walk model in which $\eta$ independent
walkers must arrive at the same surface point in order for growth to occur at
that point. These ``$\eta$-models" were originally introduced by Niemeyer et al. as
models for dielectric breakdown, and represent a useful formal extension of
DLA\cite{pietronero}.

These models were used in an important recent work of Hastings to propose
a systematic perturbative approach to DLA\cite{hastings}. This work argued that the
fractal dimension $D$ of $\eta$-model clusters collapses to $D=1$ in
spatial dimensionality $d=2$ for $\eta \ge 4$, and that this value of
$\eta_c = 4$ therefore represents an upper critical $\eta$ for these models.
Dimensions and other properties of models for $\eta < 4$ can then be
determined by perturbative renormalization in $4-\eta$. The
case of DLA ($\eta =1$) is in principle accessible, although satisfactory agreement
with the numerical result $D=1.71$ may be difficult to achieve given the large value
of
$4-\eta$ required. However, considerable computational 
difficulties arose in implementing this program. Nevertheless, rough numerical
results for the first order correction to $D$ were obtained, which agree with the
exact result expressed in Eq.~(\ref{eq:dim}) below.

Many of the ideas used by Hastings originated in the ``branched growth
model," a phenomenological treatment of DLA proposed a number of years
ago by my collaborators and myself\cite{halseyleibig}. The purpose of the
current work is to show that the branched growth model actually allows easy
computation of perturbative terms, at least to first order in $4-\eta$. This
ease can be understood as a consequence of the branched growth model
becoming exact as one approaches $\eta = 4$. In particular, I obtain the
result that the dimension $D$ is given to first order in $4 - \eta \ge 0$
by

\eq
D = 1 + \frac{1}{2}(4-\eta) + O(4-\eta)^2 .
\label{eq:dim}
\en
I obtain as well first order expressions for the multifractal
dimensions of the growth measure.

In this work, I will first review the salient features of the branched
growth model, and I will argue that to lowest order in $4-\eta$ it
well represents the dynamics of the underlying Laplacian growth process.
I compute the dimensions of the clusters by two different means, and
show that both methods give Eq. (\ref{eq:dim}). I then derive an integral formula for
the multifractal dimensions to $O(4-\eta)$, and give a simple approximation 
to the actual values of these dimensions at this order.
Finally, I discuss prospects for extending this
computation to higher orders in $4-\eta$.

The branched growth model places a fundamental importance on the
microscopic process of tip-splitting, whereby a growing branch forks
into two growing branches. This process occurs at a microscopic scale, on
the order of the particle size or cutoff $a$ in dimension. Thus the
frequency and detailed dynamics of tip-splitting is controlled by microscopic and
presumably non-universal details of the way particles attach at or near the tip of a
growing branch. We regard tip-splitting as the fundamental stochastic
process in the model; we disregard all other forms of stochasticity such
as the ``shot noise," i.e., the purely statistical variations in the
number of particles depositing at different positions in the cluster.
The reader should note that the precise role of stochasticity in DLA has
recently been quite controversial\cite{stochastic}; although we believe that the
theory to be outlined in this work is stable against obvious additional sources
of stochasticity such as shot noise, a complete understanding of the
roles of different kinds of noise requires more systematic study.

Once a branch splits into two, we follow its additional development
by implementing a deterministic version of the $\eta$-model growth rule.
Near the tip of a linear equipotential, the electric field in two dimensions diverges
as

\eq
E(w) \sim w^{-1/2}
\en
with $w$ the distance from the tip. Thus, for $\eta > 2$ the growth
measure, which is proportional to $E^{\eta} (w)$, is entirely dominated
by growth at the tips. Since we work near $\eta =4$, we need only follow
the progress of the tips in the deterministic portions of the growth,
as well as keeping track of the generation of new tips through
(stochastic) tip-splitting.

Consider two branches emanating from the same tip-splitting event. The
masses (particle numbers) of the two branches we will write as $n_L$ for the
left-hand branch, and $n_R$ for the right-hand branch; the growth measures of
the left-hand and right-hand branches we write as $p_L$, $p_R$
respectively. Defining relative growth rate and mass parameters $x$ and
$y$ respectively by

\eq
x = \frac{p_L}{p_L + p_R},
\label{eq:defx}
\en
and
\eq
y = \frac{n_L}{n_L+n_R} \equiv \frac{n_L}{n_T},
\label{eq:defy}
\en
with $n_T = n_L + n_R$, we see by elementary manipulations that

\eq
\frac{d y}{d \log(n_T)} = x - y .
\label{eq:dydn}
\en
In principle, we can define a function $G$ of the overall cluster geometry such that

\eq
\frac{d x}{d \log(n_T)} = G .
\label{eq:dxdn}
\en
In the branched growth model, $G$ is taken to be a function of $x$ and
$y$ alone. 

Let us consider a growing fork, i.e., a branch with equal sub-branches (the tines of
the fork) immediately after tip-splitting. It is convenient to describe the fork by
the conformal map $w = F(z)$ that maps the real axis in the
$z$-plane onto the fork in the physical $w$-plane; $\vert dw/dz
\vert^{-1}$ gives the local electric field at $w$. If we choose a fork
for which the angle between the tines
is $\theta_1$, the map is given by

\eq
w=F(z) = z^{\alpha_1}(z^2 - 1)^{\alpha_2}
\en
with $\alpha_1 = \theta_1/\pi $ and $\alpha_2 = 1 - (\theta_1/2 \pi)$,
where the constant $1$ is chosen arbitrarily.  The derivative of the map
$F^{\prime} (z)$ possesses zeroes at $z _{\pm} = \pm \sqrt{\alpha_1/2}$,
which correspond to the points of the fork. We can fix $\alpha_1$ by
requiring that the points are oriented towards the maximum field, so that
the fork geometry is unchanged by the growth process; this requires that

\eq
\frac{d}{dz} \left ( \frac{F^{\prime} (z)}{z-z_{+})} \right )\vert_{z=z_+}
=
\frac{d}{dz}
\left ( \frac{F^{\prime} (z)}{z-z_{-})} \right ) \vert_{z=z_-} =0
\en
which after some algebra determines $\alpha_1 = \frac{2}{5}$ or $\theta_1 =
\frac{2}{5} \pi$ \cite{hastings,tips}.

The competition between the two growing tines of the fork is
intrinsically unstable. A simple computation shows that the eigenvalue of
the instability is \cite{hastings}

\eq
\nu = \frac{\eta}{2} - 1 .
\label{eq:eigen}
\en
Since the two tines are supposed to be created with approximately constant
probability to be found near the fixed point $(x,y) = (\frac{1}{2},\frac{1}{2})$,
the eigenvalue of the instability can be related to the probability that the branch
pair is still active (i.e., one branch has not been entirely screened by the other)
as $n_T$ grows; this probability is $P \propto n_T^{-\nu}$. Since for a main branch
of length
$r$ there will have been $O(r)$ opportunities to branch, requiring that $O(1)$
sidebranch always be active implies $\nu = D^{-1}$ \cite{halseyleibig}, which already
suggests Eq.~(\ref{eq:dim}). However, it is productive to consider this
question in more detail.

\begin{figure}
\centerline{\epsfbox{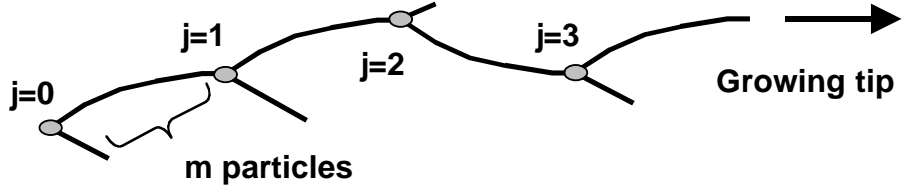}}
\medskip
\caption{The geometry of the growing branch. After each tip-splitting event, the
weaker branch is screened by the stronger. The tip-splittings are indexed by $j \ge
0$. Tip-splittings are separated by $m$ particles on the main branch.}
\label{1}
\end{figure}

Consider a growing branch of $n$ particles, which tip-splits every $m \ll
n$ particles. The side-branches thus generated persist for a certain
distance, and are then screened by the main branch. Index these
tip-splittings by $ 0 \le j \le J$ (see Figure 1). Then at each $j$ there
can be defined parameters
$x_j$, $y_j$, and $n_{j,T}$, giving the relative mass and growth
probabilities of the side-branches, as well as the total mass of the
remainder of the main branch plus the side-branch in question (the total mass to the
right of the branch point in Figure 1). We choose our definition of ``left" and
``right" in Eqs. ~(\ref{eq:defx}-\ref{eq:defy}) so that each
$x_j
\le
\frac{1}{2}$. The growth rate at the overall tip is then given by

\eq
p_{tip} = \prod_{j=0}^J(1-x_j) .
\en
Suppose that at the $j$'th branching, the initial value of $x_j$ is given
by

\eq
x_j(n_{j,T} = 1) = \frac{1}{2} - \varepsilon_j^{\nu} ,
\en
defining a random variable for the branching $\varepsilon_j$. The
distribution of $\varepsilon$, $\rho(\varepsilon)$, is chosen so that $x_j$ does not
have any singularity in its initial distribution near $x = 1/2$--this reflects the
microscopic origin of the stochasticity.

\eq
\lim_{\varepsilon \to 0} \rho(\varepsilon) = \rho_0 \varepsilon^{\nu-1} ,
\en
with the large $\varepsilon$ behavior constrained by the integrability of
$\rho(\varepsilon)$. In the branched growth model, the dependence of
$x_j$ on $n_{j,T}$ and $\varepsilon_j$ at each branching
point is the same; we now assume this to be the case in this more general
model as well. The argument below will establish that this assumption is
correct to lowest order in $4-\eta$.

If the dynamics of each branching point are the
same, then there must exist some $G(x,y)$ in Eq.~(\ref{eq:dxdn}). The values of $x$
and $y$ can then be integrated to obtain $x(h)$, $y(h)$ from Eqs.
(\ref{eq:dydn}-\ref{eq:dxdn}), with $h \propto n_T$ and $(x(0),y(0)) = (\frac{1}{2},
\frac{1}{2})$. Note that this choice of variables means that the dependence of $x$
on
$\varepsilon$ can be encoded by

\eq
x(n_T, \varepsilon) \approx x(\varepsilon n_T \equiv h) ,
\en
and similarly for $y$. This formula becomes exact for $n_T$ large. 

We can now see that to
$O(\rho_0)$

\eq
\langle p_{tip} \rangle = 1 - \sum_j \langle x_j \rangle = 1- \sum_{j=0}\int d
\varepsilon \rho_o \varepsilon^{\nu-1} x (\varepsilon n_{j,T}) ,
\en
or, extracting the dependence on $\{n_{j,T}\}$ from the integrals,

\eq
\langle p_{tip} \rangle = 1 -
\sum_j
\langle n_{j,T}^{-\nu} \rangle \left ( \rho_0 \int_0^{\infty}  dh \;
h^{\nu-1} x(h) \right ) .
\label{eq:ptip}
\en

To evaluate $\sum_{j=0}
\langle n_{j,T}^{-\nu} \rangle$, we use a simple trick. Let us first assume that

\eq
\sum_{j=0}
\langle n_{j,T}^{-\nu} \rangle = \lambda \log n .
\en
We will now confirm this form, and compute the parameter $\lambda$.
Consider the second branching at $j=1$. We have

\eq
\lambda \log n = \frac{1}{n^{\nu}} + \sum_{j=1}
\langle n_{j,T}^{-\nu} \rangle =  \frac{1}{n^{\nu}} + \langle \lambda \log 
\left [ (1-y_0) n-m \right ] \rangle ,
\en
since the number of particles in the main branch below the second sidebranching is
$n_1 = (1-y_0) n_0 - m$ (see Figure 1). Note that $n \equiv n_{0,T}$. This leads to

\eq
\frac{1}{n^{\nu}} = - \frac{\rho_0}{n^{\nu}} \lambda \int_0^{\infty} dh \;
h^{\nu-1} \log (1- y(h)) + \frac{\lambda m}{ n} \langle (1- y_0)^{-1} \rangle .
\label{eq:logy}
\en

As $h \to \infty$, the weaker branch will die at some fixed mass and $y
(h) \to \bar h / h$. Thus as $\nu \to 1$, the integral over $h$ on the right-hand
side of Eq.~(\ref{eq:logy}) diverges as

\eq
\int dh \;
h^{\nu-1} \log (1- y(h)) = -\frac{\bar h}{1-\nu} + O (1)
\en
Thus we obtain immediately

\eq
\lambda = \frac{1- \nu}{\bar h \rho_0} + O(1-\nu)^2
\en
so that the sum in Eq.~(\ref{eq:ptip}) is $O(1-\nu) = O(4-\eta)$. 

The fact that this ``propagator" sum is $\propto (1-\nu) \log n$ is the
key formal result of this work, which allows us to construct a direct
renormalization group for the dimension and other properties of the
$\eta$-models. The procedure, in principle, is as follows. First, a
naive perturbative expansion in $\rho_0$ is constructed, along the lines
of ref.~\cite{halseyexpand}. The computation above shows an example to first order
in $\rho_0$. This expansion should account both for the different contributions of
the various tips to the quantity being computed, as well as the influence of the
internal structure of the various branches on the functions $x(h)$, $y(h)$. In this
expansion, sums over
$\langle n_{j,T}^{-\nu} \rangle$ such as that appearing in Eq.~(\ref{eq:ptip}) will
appear, as well as more complex, albeit still ultimately logarithmic, sums.
Performing these sums, one will replace the original series in
$\rho_0$ with a logarithmic series in $1- \nu$. The methods of
ref.~\cite{halseyexpand} easily show that the higher order terms in $\rho_0$ will
be higher order in $1- \nu$ upon computation of these sums. This series then forms
the basis of a direct renormalization calculation of the quantity of interest.

We illustrate this by returning to the growth rate of the overall branch
tip. To zeroth order in $1-\nu$, we can represent the structure of the two
branches, which we use to compute $x(h)$ and $y(h)$, by a fork with two
growth sites at the fork tips. Thus we generalize the fork with equal-length tines to
a fork which may grow different lengths on the two sides--in so doing, there will
be a tendency for the two sides to curve away from being perfectly straight
and separated by an angle $\theta_1$, as shown in Figure 1. Standard
techniques of integrating the conformal map for Laplacian growth structures will
suffice for determining the full
$x(h)$ and
$y(h)$ for this case\cite{cmm}; below we give a simple approximation to these
quantities. To this order, there are no additional
variables describing the internal structure of the branches, so we are justified in
assuming the same $x(h)$ and $y(h)$ at each branch point. 

However, it turns out that we need not compute $x(h)$ or $y(h)$ explicitly in order
to compute the first order correction to the tip growth rate. From the definitions of
$x(h)$ and $y(h)$, we have that

\eq
\frac{d (y n_T) }{d n_T} = x ,
\en
implying by integration of parts that

\eq
\int_0^{\infty} dh \; h^{\nu-1} x(h) = (1- \nu) \int_0^{\infty} dh \;
h^{\nu-1} y(h) .
\en
The divergence of $\int dh h^{\nu-1} \log (1-y(h))$  as $\nu \to 1$ originates from
large values of $h$, or small values of $y$. Thus to lowest order in $1-\nu$ we have

\eq
\int_0^{\infty} dh
\; h^{\nu-1} \log(1-y(h)) = -\int_0^{\infty} dh
\; h^{\nu-1} y(h) ,
\en
and

\eq
\sum_j\langle n_{j,T}^{-\nu} \rangle  \rho_0 \int_0^{\infty}  dh \;
h^{\nu-1} x(h)  = \frac{ \rho_0 \int_0^{\infty}  dh \;
h^{\nu-1} x(h) }{ \rho_0 \int_0^{\infty}  dh \;
h^{\nu-1} y(h) } \log n = (1 - \nu) \log n ,
\en
so that
\eq
p_{tip} = 1 - (1- \nu) \log n + O(1-\nu)^2 .
\en
Since we expect $p_{tip} \propto n^{D^{-1} - 1}$, with $D$ mass-radius scaling
dimension
\cite{turk}, this implies with Eq.~(\ref{eq:eigen}) that

\eq
D = 1 + \frac{1}{2} (4 - \eta) + O(4-\eta)^2 ,
\en
as advertised.

To compute multifractal dimensions to $O(4-\eta)$, I use similar
techniques. Following ref.~\cite{halseyexpand}, and using $i$ as an index to
growth tips, we see that the multifractal spectrum $\sigma(q)$ for the growth measure
({\it not} the harmonic measure) is given by

\eq
\langle \sum_i p_i^q  \rangle \equiv n^{-\sigma(q)} =  1  + \lambda \log n
\left
\{
\int_0^{\infty} dh  \left [ x^q (h) + (1-x(h))^q - 1 \right ]\right \}  +
O (1- \nu)^2,
\en
yielding 
\eq
\sigma(q) = - \frac{1-\nu}{\bar h} \left \{ \int_0^{\infty} dh
\left [ x^q(h) + (1-x(h))^q - 1 \right ] \right \}  + O (1- \nu)^2 .
\label{eq:mfdim}
\en

To obtain explicit results for the multifractal dimensions, we need the trajectories
$x(h)$ and $y(h)$. I use a simple artifice that gives a useful
approximation to this trajectory. The integral in Eq.~(\ref{eq:mfdim}) is dominated
by values of
$x$ near
$x=\frac{1}{2}$. We can thus approximate the integral by taking the linear
trajectory in the $x-y$ plane near the center $(x,y) = (\frac{1}{2},\frac{1}{2})$
and extending it to the boundaries $x=0,1$. (This approximation is referred to as
``model Z" in ref.~\cite{halseyleibig}.) Explicitly, we write, taking the lowest
order $\nu = 1$,

\eq
x(h) = \cases{\frac{1}{2} - h, & if $h < \frac{1}{2}$\cr 0, & if $h \ge
\frac{1}{2} ,$\cr}
\en
and
\eq
y(h) = \cases{\frac{1}{2} - \frac{1}{2} h, & if $h < \frac{1}{2}$,\cr 1/8h ,
&if
$h
\ge
\frac{1}{2}$ .\cr}
\en
which agrees with Eq.~(\ref{eq:dydn}).

The approximate result for the multifractal dimensions is then

\eq
\sigma(q) = 2 (4-\eta) \frac{q-1}{q+1} + O(4-\eta)^2 .
\en
Note that due to our total suppression of non-growing portions of the measure, we do
not recover the identity $\sigma(0) = - D$.

The extension of these results to higher orders in $4-\eta$, and in particular to
the case of DLA ($\eta = 1$), will require some further formal development.
Reference \cite{halseyexpand} successfully computes the most divergent and next-most
divergent terms at all orders in $\rho_0$ for the multifractal dimensions for the
branched growth model; the behavior of the higher order logarithms in this case does
allow resummation of the theory for, e.g., quenched and annealed multifractal
dimensions. In our case, we need to add a family of terms representing the
deviations from the perfect branched growth model behavior, which arise from
fluctuations in the internal structure of the branches. Fortunately, there are
indications that these fluctuations are also renormalizable \cite{halseyleibig}. 

I am grateful to M.B. Hastings for drawing my attention to ref.~\cite{hastings},
and for a critical reading of this manuscript.

\end{document}